\documentclass[aps,prd]{revtex4}

\usepackage{amsmath, amsthm, amssymb, mathrsfs}
\usepackage{bm}
\usepackage{verbatim} 

\begin{document}

\title{Dark Energy, Background Independent Quantum Mechanics\\
 and the Origin of Cosmological Constant}
\author{Aalok} 
\email{aalok@uniraj.ernet.in}

\affiliation{Department of Physics, University of Rajasthan, Jaipur 302004 India;\\
 and Jaipur Engineering College and Research Centre (JECRC) 303905 India}

\date{\today}


\begin{abstract}
We explore the extended framework of the generalized quantum mechanics and discuss various aspects of neighborhood in the
construction of space in search of origin of cosmological constant. We propose to expand definition of the volume
of the phase space in eight dimensions with an overall constraint in the form of uncertainty relation as:
$(\Delta p_{x}\Delta p_{y}\Delta p_{z}\Delta E)(\Delta x\Delta y\Delta z\Delta t)\sim h^{4}$. 
We argue that the phase space volume in the eight dimensions is an appropriate representation that it should be, and
the relation $(\Delta \Lambda)(\Delta V)\sim h$ again brings it down to the reduced phase space. 
  
\end{abstract}

\maketitle

PACS number(s): 04.60.-m, 11.25.Yb\\

We explore the extended framework of the generalized quantum mechanics in search of origin of cosmological constant in the
light of the recent work [1] in geometric quantum mechanics. The construction of the quantized space, which is key to the 
understanding of cosmological constant, has been explored intensively in the context of geometric quantum mechanics [1-3].
We discuss the idea of neighborhood in the space-time in a holistic view incorporating the general settings of geometric
quantum mechanics [1-8] from various view points. The probabilistic (statistical) interpretation of QM is hidden in the
metric properties of $\mathscr{P(H)}$, and the unitary time evolution is related to the metrical structure [1, 3-8].\\ 
The distance on the projective Hilbert space is defined in terms of metric, called the metric of the
ray space [1-4, 6-10] or the projective Hilbert space $\mathscr{P}$, given by the following expression in Dirac's
notation:
\begin{equation}
ds^{2}=[\langle d\psi\mid d\psi\rangle-\langle d\psi\mid\psi\rangle\langle\psi\mid d\psi\rangle]
\end{equation}
valid for an infinite dimensional $\mathscr{H}$, has been shown to possess metric components $g_{\mu\nu}$ identified 
in terms of Compton wavelength as:
\begin{equation}
[\langle\partial_{\mu}\psi\mid\partial_{\nu}\psi\rangle-
\langle\partial_{\mu}\psi\mid\psi\rangle\langle\psi\mid\partial_{\nu}\psi\rangle]
=\frac{1}{\lambdabar_{C}^{2}}(=\frac{m_{0}^{2}c^{2}}{\hbar^{2}}).
\end{equation}

The metric in the ray space being treated by physicists as the background independent and space-time independent 
structure, can play an important role in the construction of a potential ''theory of quantum gravity''. The demand
of background independence in quantum theory of gravity calls for an extension of standard geometric quantum
mechanics [1, 3-5]. The metric structure in the projective Hilbert space is treated as background independent and
space-time independent geometric structure. It is an important insight, which can be springboard for our proposed
background independent generalization of standard quantum mechanics. For a generalized coherent state, the FS metric
reduces to the metric on the corresponding group manifold [2]. Thus, in the wake of ongoing work in the field of quantum
geometric formulation, the work in the present discussion may prove to be very useful. The probabilistic (statistical)
interpretation of QM is hidden in the metric properties of $\mathscr{P(H)}$. The unitary time evolution is also in a way
related to the metrical structure [4, 5] with Schr$\ddot{o}$dinger's equation in the guise of a geodesic equation 
on $CP(N)$. The time parameter of the evolution equation can be related to the quantum metric \textit{via}:          
\begin{equation}
(\Delta E)^{2}\equiv\langle\psi\mid H^2\mid\psi\rangle-\langle\psi\mid H\mid\psi\rangle^2;
\end{equation}
with $\hbar ds=\Delta Edt$.\\
And the Schr$\ddot{o}$dinger equation can be viewed as a geodesic equation on $CP(N)=\frac{U(N+1)}{U(N)\times U(1)}$ as:
\begin{equation}
\frac{du^a}{ds}+\Gamma_{bc}^{a}u^{b}u^{c}=\frac{1}{2\Delta E}Tr(HF_{b}^{a})u^{b}.
\end{equation}
Here $u^{a}=\frac{dz^{a}}{ds}$ where $z^{a}$ denote the complex coordinates on $CP(N)$, $\Gamma_{bc}^{a}$ is the connection
obtained from the Fubini-Study metric, and $F_{ab}$ is the canonical curvature 2-form valued in the holonomy gague group
$U(N)\times U(1)$. Here, Hilbert space is $N+1$ dimensional and the projective Hilbert space has dimenssions $N$.\\
If the metric of quantum states is defined with the complex coordinates in the quantum state space, it is known as Fubini-
Study metric which lies on the K$\ddot{a}$hler manifold or $CP(N)$ and is identified with the quotient set 
$\frac{U(N+1)}{U(N)\times U(1)}$.\\
Alternatively, the Grassmannian:
\begin{equation}
Gr(C^{N+1})=\frac{Diff(C^{N+1})}{Diff(C^{N+1}, C^{N}\times {0})},
\end{equation}
is also found to be the most appropriate representation of this symmetry preserving the required almost complex structure
[3, 8]. By the correspondence principle, the generalized quantum geometry must locally recover the canonical quantum 
theory encapsulated in $\mathbb {P^{N}}$ and also allows for mutually compatible metric and symplectic structure, supplies
the framework for the dynamical extension of the canonical quantum theory. The Grassmannian is gauged version of complex
projective space, which is the geometric realization of quantum mechanics. The utility of this formalism is that gravity 
embeds into quantum mechanics with the requirement that the kinematical structure must remain compatible with the
generalized dynamical structure under deformation [10]. The quantum symplectic and  metric structure, and therefore the 
almost complex structure, are themselves fully dynamical. Time the evolution parameter in the generalized
Schr$\ddot{o}$dinger equation, is yet not global and is given in terms of the invariant distance. The basic point as
threshold of the BIQM is to notice that the evolution equation (the generalized Schr$\ddot{o}$dinger equation) as a
geodesic equation can be derived from an Einstein-like equation with the energy-momentum tensor determined by the
holonomic non-abelian field strength $F_{ab}$ of the $Diff(\infty-1, C)\times Diff(1, C)$ type and the interpretation of
the Hamiltonian as a charge. Such an extrapolation is logical, since $CP(N)$ is an Einstein space, and its metric obeys 
Einstein's equation with a positive cosmological constant given by:
\begin{equation}
R_{ab}-\frac{1}{2}Rg_{ab}-\Lambda g_{ab}=0.
\end{equation}
The diffeomorphism invariance of the new phase space suggests the following dynamical scheme for the BIQM as:
\begin{equation}
R_{ab}-\frac{1}{2}Rg_{ab}-\Lambda g_{ab}=T_{ab}.
\end{equation}
 Moreover, the requirement of diffeomorphism invariance places stringent constraints on the quantum geometry such as the
condition of an almost complex structure (nonintegrable) on the generalized space of quantum events. This
extended framework readily implies that the wave-functions labeling the relevant space are themselves irrelevant.
They are as meaningless as coordinates in General Relativity.\\
It is fundamental issue of Physics, as the value of the cosmological constant is tied to vacuum energy density. On
the other hand, the cosmological tells us something about the large scale behaviour of the universe, since a small
cosmological constant implies the observable increase is big and (nearly) flat. Thus, the cosmological constant
relates the properties of the microscopic Physics of the vacuum to the long distance Physics on the cosmic scale (for
reviews, see ref: [11-14]).\\

We know that cosmological constant is the variance in the vacuum energy about zero mean. The variance $\Delta E$ as it 
appeared in one of the original propositions [8] of the metric of quantum states 
\begin{equation}
ds^{2}=\frac{(\Delta E)^{2}}{\hbar^{2}}dt^{2},
\end{equation}
leads to a natural question: What this uncertainty of energy stands for? It could be the variance in the vacuum energy
too. If the quantum state under consideration is the state of vacuum then
\begin{equation}
(\Delta E)^{2}=\langle 0\mid H^{2}\mid 0\rangle-\langle 0\mid H\mid 0\rangle^{2}.
\end{equation}
 It is interesting to note that there is something physical in the right hand side of equation (8) which appears as a 
geometrical form in the left hand side of the equation. The invariant $ds$ in the metric structure of quantum states is
not a distance in the dimensional sense, it is neighborhood in the topological sense. In fact, the expression 
of metric in equation (8) has also been derived [8] by taking Taylor's expansion of the quantum state $\mid \psi(t+dt)\rangle$ with time evolution and thus exploring all possible neighborhood. We know that the Taylor's expansion is a powerful
tool to examine the neighborhood of any mathematical function. It is the infinitesimally small neighborhood implied by
this expression which fills the space. 
 This expression of metric of quantum states as it appeared in one of its original propositions [8] was later
generalized in the quantum state space.
As suggested by T. W. Kibble [7] in the context of proposed generalization of quantum mechanics that the states that are
in a sense defined near vacuum can be represented by vectors in the tangent space $T_{\nu}$, and that on $T_{\nu}$ one has
all the usual structure of linear quantum mechanics expressed in the local coordinates. However, we need to specify what
is meant by ''nearness'' to the vacuum. A state $u$ is near the vacuum if the expectation value
$\langle\mid\mid\rangle_{u}$ is everywhere small [7]. Here we find a clue. The compton's wavelength in equation (2) is not
constant, while the Planck's length or $(\lambdabar_{C})_{Planck Scale}$ (say for the state of vacuum) is certainly unique
and the least [1]. At each point on the space-time manifold, the space is locally flat. Also, the vacuum (in the form of
voids in the space) today is not the same as it was in the early Universe. Locally, the vacuum energy is fixed by the
quantum theory in the tangent space, which is also the case in the Matrix theory [3]. Gauging QM generically
breaks Super-Symmetry. We do not have globally defined super-charges in space-time in the corespondance limit. This also
explains- why there is cosmological constant [10].\\
One important element of this approach to quantum gravity is the existence of correspondence limit between the dynamical
quantum theory and the Einstein's classical theory of gravity coupled to matter. At long wavelengths, once we map the
configuration space to space-time, we have General Relativity. Turning off dynamics in the quantum configuration space
recovers the canonical quantum mechanics [1, 3-5].\\ 
Space-time is locally indistinguishable from flat space (zero cosmological constant). Thus, instead of working with the 
space-time manifold, we ought to employ a larger geometric structure whose tangent spaces are the canonical Hilbert spaces
of consistent quantum mechanics of gravitons [2]. The equivalence principle we employ, relies on the universality and
consistency of quantum mechanics at each point. In every small local neighbourhood at this larger structure, the notion
of quantum mechanical measurement is identical.\\

The observed value of the cosmological constant has a natural interpretation as fluctuations about the zero mean.
The observable smallness of the cosmological constant should tell us something fundamental
about the underlying microscopic nature. To explore the reasons, we analyze the following [2, 9, 10] quantized relation: 
\begin{equation}
(\Delta \Lambda) (\Delta V)\sim h.
\end{equation}
Here, the space-time volume and the cosmological constant should be regarded as conjugate quantities and they fluctuate 
accordingly in the quantum theory. The canonical quantum expectation value of the cosmological constant vanishes. What
is meant (and observed) by vacuum energy is the fluctuation in $\Lambda$. Consequently, one can relate the smallness of
the observed cosmological constant to the largeness of space-time.\\
A manifold is constructed out of an atlas of coordinate charts. An infinitesimally small neighbourhood about any point 
is flat. The small cosmological constant could be regarded as the consequence of patching together the Physics of 
locally flat spaces consistent with the existence of canonical gravitational quanta.\\
In string theory, semi-classically, the space-time is  $\mathscr {M}_{4}\times \mathscr{K}_{6}$, where
$\mathscr {M}_{4}$ is observed macrospace-time, and $\mathscr {K}_{6}$ is the compact space, such as Calabi-Yau manifold.
And the smallness of the observed cosmological constant is a statement about the largeness of the manifold
$\mathscr{M}_{4}$. As it is the product that appears, and should be regarded as canonically conjugate quantities.
In a quantum theory, we expect that the fluctuation in one obvservable related to fluctuations in its conjugate, such as:
$(\Delta \Lambda)(\Delta V)\sim h$. In fact it is energy-time uncertainty relation in the space-time (the string theory
target space). The preferred value of the cosmological constant is certainly zero [2, 9, 10]. The existence of a measured
vacuum energy is the consequence of quantum fluctuations about the zero value. The fluctuations in $\Lambda$ are inversely
related to the fluctuations in the volume $V$.\\
In a semiclassicaltheory of gravity, the cosmological constant arises in the Einstien-Hilbert action as a prefactor for
the volume of the four-dimensional [2] space-time $\mathscr{M}_{4}$:
\begin{equation}
V=\int{{d}^{4}x \sqrt{-g}}.
\end{equation}
It is not surprising that the metric of quantum state space has its definition in statistical mechanics also [2, 14, 15],
and is alternatively expressed as statistical distance on the space of quantum events uniquely determined by the size of 
statistical fluctuations occurring in measurements performed to tell one event from another. This distance between two
statistical events is given in terms of number of distinguishable events, thus forming the space with the associated
Riemannian metric $ds^{2}\equiv\sum_{i} \frac{dp_{i}^{2}}{p_{i}}=\sum dX_{i}^{2}$, where $p_{i}\equiv X_{i}^{2}$ denote
individual probabilities. The distance in the probability space is nothing but the celebrated Fisher distance of the
information theory and can be written as [15]:
\begin{equation}
ds_{12}=cos^{-1}(\sum_{i}\sqrt {p_{1i}}\sqrt {p_{2i}}).
\end{equation}
Within a quantum theory, events cannot be localized to arbitrary precision. Only for high-energies does it even make sense
to speak of a local region in the space-time where an interaction takes place. This is simple consequence of the 
energy-time uncertainty relation. Fluctuations in the volume of space-time are fixed by statistical fluctuations in the
number of degrees of freedom of the gauged quantum mechanics. To enumerate the degrees of freedom, we employ the 
statistics of distinguishable particles. The fluctuation is given by a Poisson's distribution, which is typical for
coherent states.\\
Thus, the studies of cosmological constatnt, or to say studies of space-time by means of statistical mechanics have come
a long way. And, we see further possibilities of break-through in the understanding of cosmological constant by means of
statistical mechanics in the following discussion.\\
The fluctuations of relevance for us lie in the number of Planck sized cells that fill up the configuration space
(the space in which quantum events transpire). The uncertainty principle prevents us from representing a moving
physical object by a single vector. This is because such a representation would amount to specifying both the position
and the momentum exactly. Thus, phase space is divided into cells with volume:
\begin{equation}
(\Delta p_{x}\Delta p_{y}\Delta p_{z})(\Delta x\Delta y\Delta z)\sim h^{3}
\end{equation}
Equivalently, one can say that the state of a system cannot be specified more closely than by saying that the tip of the
vector representing it lies in one of these cells.\\
The volume of the phase space by equation (13) is with the consideration that the energy $E$ of each phase space cell is
fixed.
Now, we propose to expand the volume of the phase space in eight dimensions, ensuring the underlying formalism to be
manifestly covariant. This is with the consideration that phase space cells also observe fluctuations in the energy as 
$E+\Delta E$ or $E-\Delta E$. Also, we emphasize the need to widen the covariant formalism in thermodynamics and
statistical mechanics for the sake of generalizations. The extended phase space in eight dimensions is thus natuarally
associated with the constraint of uncertainty as:
\begin{equation}
(\Delta p_{x}\Delta p_{y}\Delta p_{z}\frac {\Delta E}{c})(\Delta x\Delta y\Delta z c\Delta t)\sim h^{4},
\end{equation}
or     
\begin{equation}
(\Delta p_{x}\Delta p_{y}\Delta p_{z}\frac {\Delta E}{c})\Delta V\sim h^{4}.
\end{equation}
Thus, using the relation 
$(\Delta \Lambda)(\Delta V)\sim h$
we again reduce the phase space to the following relation:
\begin{equation}
(\Delta p_{x}\Delta p_{y}\Delta p_{z}\frac {\Delta E}{c})(\frac {1}{\Delta \Lambda})\sim h^{3}.
\end{equation} 
From this, we can also conclude
\begin{equation}
(\Delta \Lambda)\sim (\frac{\Delta p_{x}\Delta p_{y}\Delta p_{z}\Delta E}{ch^{3}}).
\end{equation}
Now we argue that the phase space volume in the eight dimensions is an appropriate representation that it should be,
and the relation $(\Delta \Lambda)(\Delta V)\sim h$ again brings it down to the reduced phase space given by
equation (13).\\
The expansion of the universe is observed to be the driving factor that affects the value of cosmological constant.
Therefore the rate of expansion of the universe certainly plays a role in affecting the cosmological constant.
Consequently, it it just not possible that the rate of expansion of the universe in different directions has no effect
on the cosmological constant! It is quite apparent that the cosmological constant arises not because of the variance in
the vacuum energy alone. The variance in all the components of four momenta of vacuum phase cells gives rise to it.
But, as the ensemble (universe) on the whole is isotropic, its rate of expansion in different directions is uniform, and 
effectively the cosmological constant at large turns out to be equivalent to the variance in the vacuum energy only.
One may call it a retro-realization or a reverse approach to the realization of this truth. And this gives rise to vast 
possibilities of further investigations. 


\begin{acknowledgments}

The author wishes to thank Prof.A. Ashtekar for explaining the need and importance of background independent
quantum mechanics.
\end{acknowledgments}

\end{document}